\documentclass[aps, prl,twocolumn]{revtex4} 

\usepackage{amsmath}               
\usepackage{graphicx}

\begin{document}


\title{Anderson Localization in Nonlocal Nonlinear Media}
\author{Viola Folli$^{1,2}$, Claudio Conti$^{1,2}$}
\affiliation{
$^1$Department of Physics, University Sapienza, Piazzale Aldo Moro, 5, 00185, Rome (IT)\\
$^2$ISC-CNR, Dep. Physics, Univ. Sapienza, Piazzale Aldo Moro 5, 00185, Rome (IT)}
 
\begin{abstract}
The effect of focusing and defocusing nonlinearities on 
Anderson localization in highly nonlocal media is theoretically and numerically investigated. 
A perturbative approach is developed to solve the nonlocal nonlinear Schroedinger equation in the presence of a random potential, showing that nonlocality stabilizes Anderson states.
\end{abstract}

\maketitle
\begin{figure}[t]
\includegraphics[width=\columnwidth]{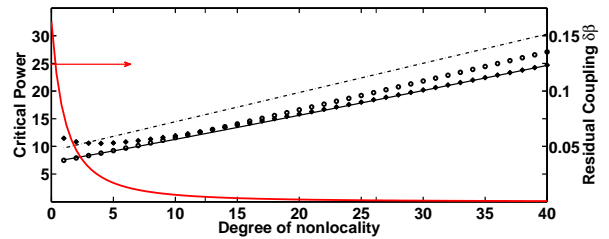}
\caption{(Color online) Left axis: critical power for exponential ($P_c=|\beta_0|(2\sigma+l)$, dashed-dotted line), Gaussian ($P_c=l |\beta_0|\exp(-\sigma^2/l^2) /Erfc(\sigma/l)$, continuous line), quadratic ($P_c=|\beta_0|/|\chi(0)|(1-l^2/2\sigma^2)$, dots) and 
rectangular ($P_c=2 \sigma|\beta_0|/(1-e^{-2\sigma/l})$, diamonds) response functions, for a single realization of the disorder; right, residual coupling $\delta\beta$ Vs degree of nonlocality, averaged over $10$ disorder realizations ($V_0=1$).}
\label{fig1} \end{figure}
\begin{figure}[t]
\includegraphics[width=\columnwidth]{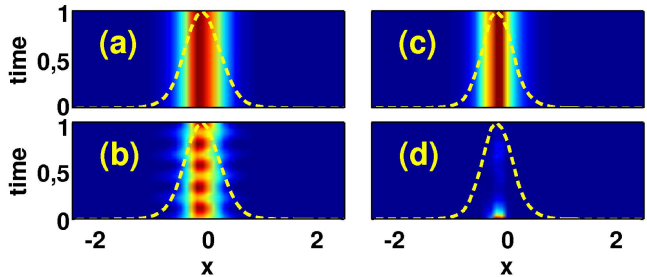}
\caption{(Color online) Evolution of the ground-state intensity for a fixed disorder realization, for $\sigma=10$, $V_0=10$; focusing (attractive) case $s=1$, for $P=0.04P_{c}$ (a) and
for $P=4P_{c}$ (b); defocusing (repulsive) case $s=-1$, for $P=0.04P_{c}$ (c) and for $P=4P_{c}$ (d). The superimposed dashed line in the panels
represents the ground-state at the initial time. 
\label{fig2}} \end{figure}
\begin{figure}[t]
\includegraphics[width=\columnwidth]{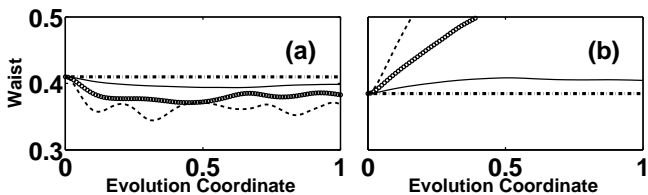}
\caption{(Color online) Wavefunction waist for focusing (a), and defocusing (b) cases, for 
$P=0.04P_{c}$ (dash-dotted line), $P=0.64P_{c}$ (continuous line), $P=2P_{c}$ (dotted line), $P=4P_{c}$ (dashed line),
($\sigma=10$, $V_0=10$). Results averaged over $10$ disorder realizations.
}
\label{fig3} \end{figure}

\noindent
Disorder and nonlinearity are two leading mechanisms 
promoting wave localization. On one hand, a sufficient strength of disorder fosters 
the transition from a diffusive regime to a
 wave-function exponentially decaying over a characteristic distance $l$;
a scenario commonly referred to as Anderson localization \cite{Anderson58,ShengBook,Wiersma97,John87,Conti08PhC, AspectNature2008, InguscioNature2008, Kivshar10, kartashov08, Swartz07}. 
On the other hand, in a nonlinear medium, diffraction, or dispersion, can be compensated by the nonlinear properties; 
when the characteristic lengths of these phenomena (nonlinearity and diffraction) are comparable, localized solitary waves, or solitons, emerge. 
It is well known that nonlocality in the nonlinear response can largely affect the localized wave-forms, depending on the degree of nonlocality $\sigma$ 
\cite{Bang02,rasmussen2005theory,skupin10, Conti03}. However the interplay between the disorder induced localization length $l$
and the characteristic length of nonlocality $\sigma$ has never been considered before.
Here we analyze the effect of a nonlocal nonlinearity on Anderson localization; we show that 
in the framework of the highly nonlocal approximation, it is possible to derive closed form expressions to describe the role of nonlinearity on Anderson localizations,
and that these states become more stable when the degree of nonlocality increases, meaning that the power needed to destabilize them increases with $\sigma/l$;
this result unveils a fundamental connection between nonlocality and disorder.
\\\noindent \textit{Model ---} The nonlocal nonlinear Schroedinger equation reads as
\begin{equation}
\label{RNLS}
i\psi_t+\psi_{xx}=V(x)\psi-s \psi \int_{-\infty}^{+\infty}\chi(x'-x)|\psi(x')|^2 dx'
\end{equation}
where $\psi=\psi(x,t)$, $V(x)$ is a random potential and $\chi(x)$ is the response function of the nonlocal medium normalized such that $\int \chi(x) dx=1$; 
 $s\pm 1$ corresponds to a focusing ($s=1$) or defocusing  ($s=-1$) nonlinearity. Eq. (\ref{RNLS}) applies to a variety of physical problems, including nonlinear optics and Bose Einstein condensation \cite{Folli10}.
We define the ``unperturbed'' Hamiltonian as 
$H_0=-\partial_{x}^2+V(x)$; its eigenstates are written as $H_0\psi_n=\beta_n \psi_n$
with $\left(\psi_n, \psi_m\right)=\delta_{nm}$.
$H_0$ sustains exponentially localized states, corresponding to negative eigenvalues $\beta_n$. 
The fundamental state can be approximated by $\psi_0(x)=\frac{1}{\sqrt{l}}e^{-|x-x_0|/l}$\text{,}
where the average localization length $l$ is determined by the strength of the random potential $V_0$, and $x_0$ is the location of the eigenfunction with eigenvalue $\beta_0$. 
In the following, without loss of generality, we assume that the horizontal axis has been shifted such that $x_0=0$.
From $H_0$, one roughly finds the link between the eigenvalue and the average localization lenght $l$
by a taking the average wavefunction $\psi_0$:
$\langle\beta_0\rangle=\langle \left(\psi_0,H_0 \psi_0 \right)\rangle \cong -{1/l^2}$; the lowest energy state has the highest degree of localization.
\\\noindent\textit{Highly Nonlocal Limit ---} 
The nonlocality is described by $\chi(x)$, which is typically bell-shaped with a characteristic length $\sigma$. 
For an average localization length $l$ much shorter than the nonlocality degree $\sigma$ of the medium,
the response function $\chi(x'-x)$ can be expanded around the localization center, $x'=0$, and 
one has in (\ref{RNLS})
\begin{equation}
\psi\int_{-\infty}^{+\infty}\chi(x'-x)|\psi(x')|^2 dx'=P\chi(x)\text{.}
\end{equation}
In this {\it highly nonlocal limit} (HNL), the nonlinearity can be treated 
as an interaction Hamiltonian $H_{int}=-s\chi(x)P$,
where $P=\int_{-\infty}^{+\infty}|\psi(x)|^2 dx$ is overal energy, or
beam power, which is conserved during evolution after Eq.(\ref{RNLS}). 
This limit is valid in the regime of a wave dominated
by a single localization, which, without loss of generality, is taken centered at $x_0=0$.
This also holds true as far as during the dynamics, additional localization are generated among those located in proximity of $x=0$. Conversely, if two distant localizations are excited, $\chi(x)$ will be composed by two nonlocal responses centered in the two Anderson states, this case will be investigated elsewhere. 
The HNL allows to apply the standard perturbation theory of quantum mechanics for deriving closed form expressions for the effect of nonlinearity on the Anderson states.
We write the field as an expansion in $P$, $\psi=\sqrt{P}\left(\psi_0+ P \psi^{(1)}+ P^2 \psi^{(2)} +...\right)$,
where we take at the leading order $\psi=\sqrt{P}\psi_0$ to focus on the
effect of nonlocal nonlinearity on the fundamental state.
We obtain the correction to the Anderson ground state (at second order in $P$) eigenvalue:
\begin{equation}
\label{correction}
\beta_0(P)=\beta_0-s P\chi_{00}+P^2\sum_{n\neq0}\frac{|\chi_{n0}|^2}{\beta_0-\beta_n}\text{,}
\end{equation}
with the matrix elements of the nonlocality given by
$\chi_{nm}=\int \chi(x) \psi_n(x) \psi_m(x) dx$.
\noindent As the degree of nonlocality increases,
$\chi(x)$ can be treated as a constant in $\chi_{nm}$,
such that $\chi_{mn}=\chi(0)\delta_{nm}$.
This shows that the perturbation to the Hamiltonian is diagonal, 
therefore, in the HNL, the effect of the
nonlocality is to shift the eigenvalue such that $\beta_0(P)=\beta_0-s \chi(0) P$, where $\chi(0)$ depends on the specific $\chi(x)$, and explicitly contains the degree of nonlocality.
\noindent The overall Hamiltonian $H=H_0+H_{int}$ 
is diagonal in the same states of $H_0$, hence Anderson localizations
turn out to be eigenstates also in the presence of the nonlocality. 
These localizations are hence expected to be extremely robust with respect to the nonlinearity.  
\\{\it Instability of Anderson states ---}
The effect of nonlinearity on Anderson states becomes relevant
when the term linear in $P$ is comparable with $\beta_0$ (higher order corrections vanish in the HNL);
this allows to define through Eq.~(\ref{correction}), 
the critical power $P_c$ :
\begin{equation}
\label{critical power}
P_c=\frac{|\beta_0|}{\int_{-\infty}^{+\infty}|\psi_0(x)|^2\chi(x) dx}.
\end{equation}
For a defocusing medium ($s=-1$), this is the power needed to change the sign of 
the eigenvalue $\beta_0(P)$, from negative to positive;
such that the localization is destroyed.
Conversely, for a focusing medium ($s=1$), this can be interpreted as the
power were the average degree of localization is strongly affected by
nonlinearity, indeed as $l(P)\cong 1/\sqrt{\langle \beta(P)\rangle}$, one has
\begin{equation}
l(P)\cong \frac{l}{\sqrt{1+s P/\langle P_c\rangle}}
\label{locP}
\end{equation}
such that at critical power, the localization length is reduced
by a factor $\sqrt{2}$ for the focusing case $s=1$, and 
diverges for the defocusing case $s=-1$. This shows that for $P>P_c$ no localized states are expected for $s=-1$, as
the corresponding eigenvalue changes sign.
This trends applies as far as additional effects, like
the excitation of further localizations, occur.
From Eq.~(\ref{critical power}), we obtain the expression for $P_c$ in the HNL, 
$P_c\equiv \hat{P}_c=|\beta_0|/|\chi(0)|$ with $\langle\hat{P}_c\rangle\cong \sigma/2 l^2$ [$\chi(0)=1/2\sigma$ for an exponential nonlocality].
Because of the Cauchy-Schwarz inequality, one has 
$\int_{-\infty}^{+\infty}|\psi_0(x)|^2\chi(x) dx<1$, and one
readily sees that for a finite nonlocality $P_c<\hat{P}_c$; 
as the nonlocality increases the power needed to destabilize the Anderson states grows.
\\A useful measure to quantify the effect of a nonlocal nonlinearity on Anderson states is the residual value of $\beta(P)$ at the critical power $P_c$, which can be written as
\begin{equation}
\delta\beta(\sigma)=\frac{\beta_0(P_c)}{\beta_0}=\sum_{n\neq 0}\frac{|\chi_{n0}|^2}{|\chi_{00}|^2(1-\beta_n/\beta_0)} \text{.}
\end{equation}
As it is determined by the off-diagonal elements $\chi_{n0}$, $\delta\beta$ can be taken as the ``residual coupling'' due to the nonlocal nonlinearity, which vanishes in the HNL limit.
In Fig.\ref{fig1}, we numerically show [for an exponential $\chi(x)$] that $\delta\beta$ goes to zero when increasing $\sigma$: as the nonlocality increases 
the nonlinear coupling of $\psi_0$ with other states is moderated, hence it tends to behave as an eigenstate of the system even if nonlinearity is present 
(however, its degree of localization may be largely affected).
\\\noindent \textit{Nonlocal Responses ---}
We analyzed a few specific case of response function $\chi(x)$ \cite{Bang02} (rectangular $\chi=1/(2\sigma)$ for $|x|<\sigma$, $\chi=0$ elsewhere; exponential $\chi(x)=e^{-|x|/\sigma}/(2\sigma)$; Gaussian $\chi(x)=e^{-x^2/\sigma^2}/\sqrt{\pi \sigma^2}$; and quadratic $\chi(x)=\chi(0)+\chi_2 x^2$). In Fig.~\ref{fig1}, we report the behavior of the critical power as a function of the nonlocality degree $\sigma$ for the analyzed response functions.
In all of these cases, the calculated critical power is linearly dependent on the unperturbed eigenvalue of the state. 
So, the higher the strength of the disorder $V_0$, the lower $\beta_0$, 
the higher the power needed to affect the localization. 
Furthermore, we emphasize that the critical power, depending on the Anderson eigenvalue,
has a statistical distribution depending on the disorder configuration. 
\\\textit{Numerical Results ---} We numerically solved Eq.(\ref{RNLS}) for fixed disorder
configurations: we consider a Gaussianly distributed $V(x)$ with zero mean and standard deviation $V_0$, 
first obtain the eigenstates, then by using a pseudo-spectral Runge-Kutta algorithm, 
we evolve the Anderson localizations in a 
nonlocal medium with a given $\chi(x)$ [an exponential response hereafter, similar results are obtained for other $\chi(x)$]. Fig.~\ref{fig2} shows the dynamics of the ground-state intensity: for $P<P_c$ the state remains almost unperturbed (Fig.~\ref{fig2}(a), attractive case; Fig.~\ref{fig2}(c), repulsive case). By increasing the power beyond the critical threshold, we observe two different phenomena. 
In the focusing case, the state becomes more localized Fig.~\ref{fig2}(b). At higher powers, a temporal beating pattern can be observed. This is mainly due to the coupling of the Anderson ground state with other localizations. In fact, for $\sigma=10$ the highly nonlocal limit is not satisfied, and a residual nonlinear coupling between the Anderson ground state 
and additional localized modes is present (see Fig.~\ref{fig1}, 2nd y-axis), and causes the observed oscillations.
In the defocusing case, we observe the breaking of the Anderson localization, as expected (Fig.~\ref{fig2}(d)).
Fig.~\ref{fig3} shows the localization length of the ground state for various powers.
In the focusing case, the localization length decreases and the beating pattern is observed. For $s=-1$, the perturbation delocalizes the eigenmode as $P>P_c$.
\\ {\it Conclusions ---} 
We reported on a theoretical analysis of the effect of a nonlocal nonlinearity
on disorder induced localization.
We derive explicit formulas to predict the critical power for 
destabilizing the Anderson states, in quantitative agreement with numerical simulations.
We have shown that an increasing degree of nonlocality produces a substantial growth of the power needed
to destabilize Anderson states, which turn out to be very robust with respect to nonlinear effects. 
This can also be explained by the fact that nonlocality reduces the coupling between Anderson states.
These results may stimulate new experimental investigations, can be extended to several related problems, as quantum phase diffusion and coherence \cite{Peschel11}, ultrashort pulses in fibers \cite{ContiLinearons10}, second harmonic generation \cite{Conti10} and Bose-Einstein condensation \cite{AspectNature2008,InguscioNature2008},
and furnish a novel theoretical framework for the interplay of nonlinearity and disorder
\cite{ContiLeuzziPRB2011}.
\\We acknowledge CINECA-ISCRA and the Humboldt foundation.
The research leading to these results has received funding from the
European Research Council under the European Community's Seventh Framework Program 
(FP7/2007-2013)/ERC grant agreement n.201766.


\begin{thebibliography}{10}
\newcommand{\enquote}[1]{``#1''}

\bibitem{Anderson58}
P.~W. Anderson, Phys. Rev. \textbf{109}, 1492 (1958).

\bibitem{ShengBook}
P.~Sheng, ed., \emph{Scattering and Localization of Classical Waves in Random
  Media} (World Scientific, Singapore, 1990).

\bibitem{Wiersma97}
D.~S. Wiersma, P.~Bartolini, A.~Lagendijk, and R.~Righini, Nature \textbf{390},
  671 (1997).

\bibitem{John87}
S.~John, \prl \textbf{58}, 2486 (1987).

\bibitem{Conti08PhC}
C.~Conti and A.~Fratalocchi, Nat. Physics \textbf{4}, 794 (2008).

\bibitem{AspectNature2008}
J.~Billy, V.~Josse, Z.~Zuo, A.~Bernard, B.~Hambrecht, P.~Lugan, D.~Clement,
  L.~Sanchez-Palencia, P.~Bouyer, and A.~Aspect, Nature \textbf{453}, 891
  (2008).

\bibitem{InguscioNature2008}
G.~Roati, C.~D'Errico, L.~Fallani, M.~Fattori, C.~Fort, M.~Zaccanti,
  G.~Modugno, M.~Modugno, and M.~Inguscio, Nature \textbf{453}, 895 (2008).

\bibitem{Kivshar10}
I.~V. Shadrivov, K.~Y. Bliokh, Y.~P. Bliokh, V.~Freilikher, and Y.~S. Kivshar,
  Phys. Rev. Lett. \textbf{104}, 123902 (2010).

\bibitem{kartashov08}
Y.~V. Kartashov, V.~A. Vysloukh, and L.~Torner, Phys. Rev. A \textbf{77},
  051802 (2008).

\bibitem{Swartz07}
T.~Schwartz, G.~Bartal, S.~Fishman, and M.~Segev, Nature \textbf{446}, 52
  (2007).

\bibitem{Folli11}
V.~Folli and C.~Conti, Opt. Lett. \textbf{36}, 2830 (2011).

\bibitem{Bang02}
O.~Bang, W.~Krolikowski, J.~Wyller, and J.~J. Rasmussen, \pre \textbf{66},
  046619 (2002).

\bibitem{rasmussen2005theory}
P.~Rasmussen, O.~Bang, and W.~Kr{\'o}likowski, Physical Review E \textbf{72},
  066611 (2005).

\bibitem{skupin10}
F.~{Maucher}, W.~{Krolikowski}, and S.~{Skupin}, arXiv:1008.1891 (2010).

\bibitem{Conti03}
C.~Conti, M.~Peccianti, and G.~Assanto, \prl \textbf{91}, 073901 (2003).

\bibitem{Folli10}
V.~Folli and C.~Conti, Phys. Rev. Lett. \textbf{104}, 193901 (2010).

\bibitem{Peschel11}
S.~Batz and U.~Peschel, Phys. Rev. A \textbf{83}, 033826 (2011).

\bibitem{ContiLinearons10}
C.~Conti, M.~A. Schmidt, P.~S.~J. Russell, and F.~Biancalana, Phys. Rev. Lett.
  \textbf{105}, 263902 (2010).

\bibitem{Conti10}
C.~Conti, E.~D'Asaro, S.~Stivala, A.~Busacca, and G.~Assanto, Opt. Lett.
  \textbf{35}, 3760 (2010).

\bibitem{ContiLeuzziPRB2011}
C.~Conti and L.~Leuzzi, Phys. Rev. B \textbf{83}, 134204 (2011).

\end{thebibliography}
\end{document}